\documentclass{icrc2009}

\usepackage{graphicx}   
\usepackage[caption=false]{caption}    
\usepackage[font=footnotesize]{subfig} 
\usepackage{fixltx2e}
\usepackage{url}

\newcommand{\shorttitle}[1]%
{\markboth{Proceedings of the 31\MakeLowercase{$^{st}$} ICRC, {\L}\'{o}d\'{z} 2009}{#1} }
\newcommand{\etal}{\MakeLowercase{\textit{et al. }}} 

\begin{document}
\title{MAGIC observations of Mkn 421 in 2008, and related optical/X-ray/TeV MWL study}


\author{\IEEEauthorblockN{Giacomo Bonnoli\IEEEauthorrefmark{1},
    Ching Cheng Hsu\IEEEauthorrefmark{2},
Florian Goebel\IEEEauthorrefmark{3},
Elina Lindfors \IEEEauthorrefmark{4},
Pratik Majumdar\IEEEauthorrefmark{5},\\
Konstancja Satalecka\IEEEauthorrefmark{5},
Antonio Stamerra\IEEEauthorrefmark{1},
Fabrizio Tavecchio\IEEEauthorrefmark{6}, and
Robert Wagner\IEEEauthorrefmark{2},\\ for\\ the MAGIC Collaboration}
\IEEEauthorblockA{\IEEEauthorrefmark{1}Dipartimento di Fisica, Universit\`{a} degli Studi di Siena, Via Roma 56, 53100 Siena, Italy}
\IEEEauthorblockA{\IEEEauthorrefmark{2}Max-Planck-Institut fur Physik, F\"{o}hringer Ring 6, 80805 M\"{u}nchen, Germany}
\IEEEauthorblockA{\IEEEauthorrefmark{3}deceased}

\IEEEauthorblockA{\IEEEauthorrefmark{4}Tuorla Observatory, Department of Physics ad Astronomy, University of Turku, FI-20014 Turku, Finland}
\IEEEauthorblockA{\IEEEauthorrefmark{5}DESY, Platanenallee 6, 15738 Zeuthen, Germany}
\IEEEauthorblockA{\IEEEauthorrefmark{6}INAF, Osservatorio Astronomico di Brera, via E. Bianchi 46, I-23807 Merate, Italy}
}

\shorttitle{Bonnoli, G. \etal MAGIC Mkn 421 MWL observations in 2008}
\maketitle

\begin{abstract}
 The HBL-type blazar Markarian 421 is one of the brightest TeV gamma-ray sources of the Northern sky. From December 2007 until June 2008 it was intensively observed in the VHE ($E>100\,$ GeV) band by the MAGIC gamma-ray telescope.
The source showed intense and prolonged activity during the whole period. In some nights the integral flux rose up to 3.6 Crab units ($E>200\,$ GeV). 
Intra-night rapid flux variations were observed. 

We compared the optical (KVA) and X-ray (RXTE-ASM, Swift-XRT) data with the MAGIC VHE data, investigating the correlations between different energy bands.

  \end{abstract}

 
\section{Introduction}

Mkn 421, is one of the closest ($z=0.031$, \cite{1991trcb.book.....D}) and brightest extragalactic TeV sources; therefore the first detected \cite{1992Natur.358..477P} and one of the best studied. 

It belongs to the High-peaked  \emph{blazar} (HBL) class. According to the widely accepted unified AGN model \cite{1995PASP..107..803U}, this means that the optically detected giant elliptical galaxy hosts a radio-loud AGN, with relativistic jet closely aligned to the line of sight. In all wavebands, the radiation spectrum is clearly dominated by broad, non-thermal components, that produce the "double-bumped'' $\nu F_{\nu}$ profile which is characteristic of \emph{blazars}. Currently, the low energy bump is regarded as synchrotron emission from the relativistic electrons within the jet.  More debated is the nature of the high energy bump, for which both hadronic (e.g. \cite{1993A&A...269...67M,2003APh....18..593M}) and leptonic (e.g. \cite{1992ApJ...397L...5M}) processes are invoked by different models, but still none of these can be discarded with strong evidence. 

Moreover \emph{blazars} show intense variability in every band, on different timescales, which at GeV-TeV energies can be of few minutes \cite{2007ApJ...669..862A}.
The most important test of validity of emission models relies on how well they can account for the evolution of observational parameters of these sources, as measured in multi-band observations \cite{1998ApJ...509..608T}. For instance, correlations between variability in X-ray and VHE bands are naturally explained by leptonic models (e.g. \cite{1997jena.confE.279M}) such as External Compton (EC) or Synchrotron-Self Compton (SSC), while is more difficult to account for "orphan'' flares \cite{2004ApJ...601..151K} in these contexts\cite{2006ApJ...651..113K}.

In the case of Mkn 421 the low energy (synchrotron) bump peaks at $keV$ energies, which is typical of the HBL subclass, and can be observed with current soft X-ray telescopes, such as Swift-XRT.
The high energy component peaks at $\sim 10-100\, GeV$, allowing for observation with Imaging Air Cherenkov Telescopes (IACT). Multiwavelenght observations of this bright source, involving X-ray and VHE instruments can therefore cast deeper insight on the relevant processes in the jets. The rapid variability of the source imposes careful coordination of the observations, as time variability is so strong, especially at higher energies \cite{1996Natur.383..319G}, that the state of the source cannot be safely assumed as constant between different observation windows, especially when exceptionally active.

\section{The MAGIC Telescope}

MAGIC\footnote{Major Atmospheric Gamma-ray Imaging Telescope} is located on the western Canarian Island of La Palma, at the Observatory of Roque de Los Muchachos ($28.75^\circ$N, $18.76^\circ$W$, 2225\,$ m a.s.l.). With its tessellated parabolic mirror ($D= 17\,$m, $f/D=1$), it is the largest single dish telescope for VHE astronomy in operation. This allows for the lowest energy threshold amongst IACTs: the trigger threshold of the telescope is currently $60\,$ GeV for observations close to the zenith.

The  MAGIC camera, installed at the primary focus of the mirror, is made of 576 photomultipliers (PMT), arranged in an hexagonal shape; the observed field of view is $3.5^\circ$.
The PMT analog  signals are routed through optical fibers, then digitized by means of FADCs and stored to disk if trigger criteria are fulfilled. More details can be found in dedicated papers (e.g.\cite{2005ICRC....5..359C}).  

All the MAGIC observations  considered for the present study are subsequent to a major hardware upgrade, completed in February 2007, that enhanced (from 300 MHz to 2 GHz) the time sampling capability of the DAQ. This allowed for better rejection of the Night Sky Background (NSB) and introduction of new refined analysis techniques \cite{2008ICRC....3.1393T} based on the time properties of Cherenkov signals.

\begin{figure*}[!th]
  \centering
  \includegraphics[width=4.5in]{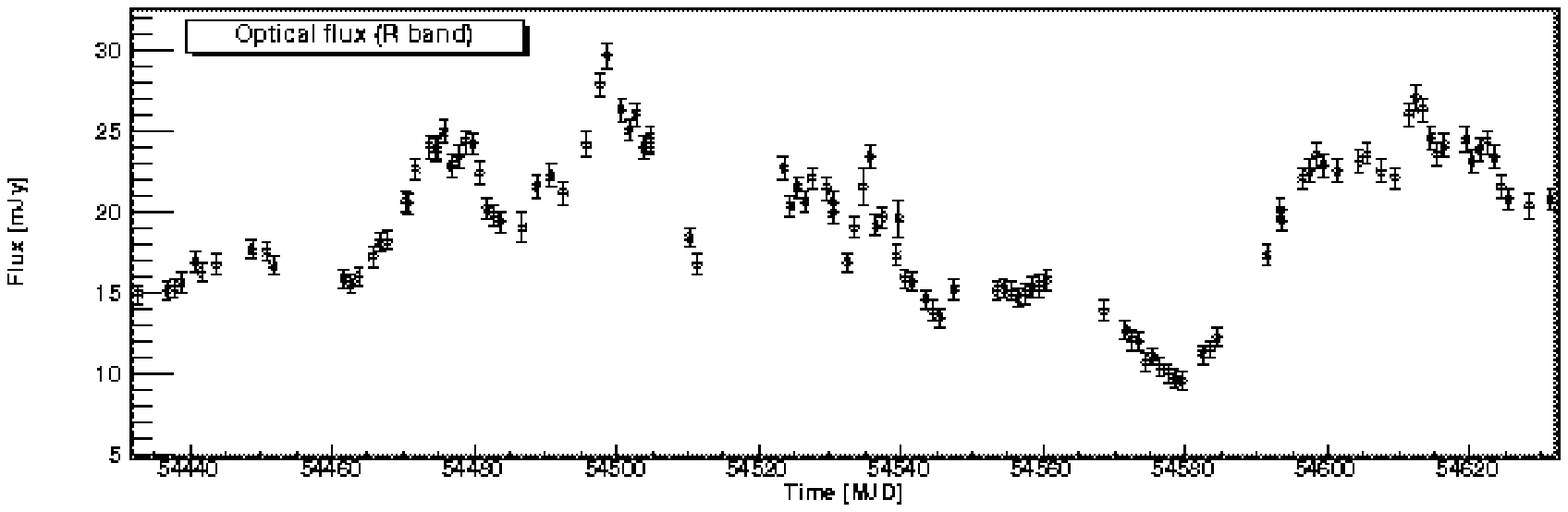}
  \includegraphics[width=4.5in]{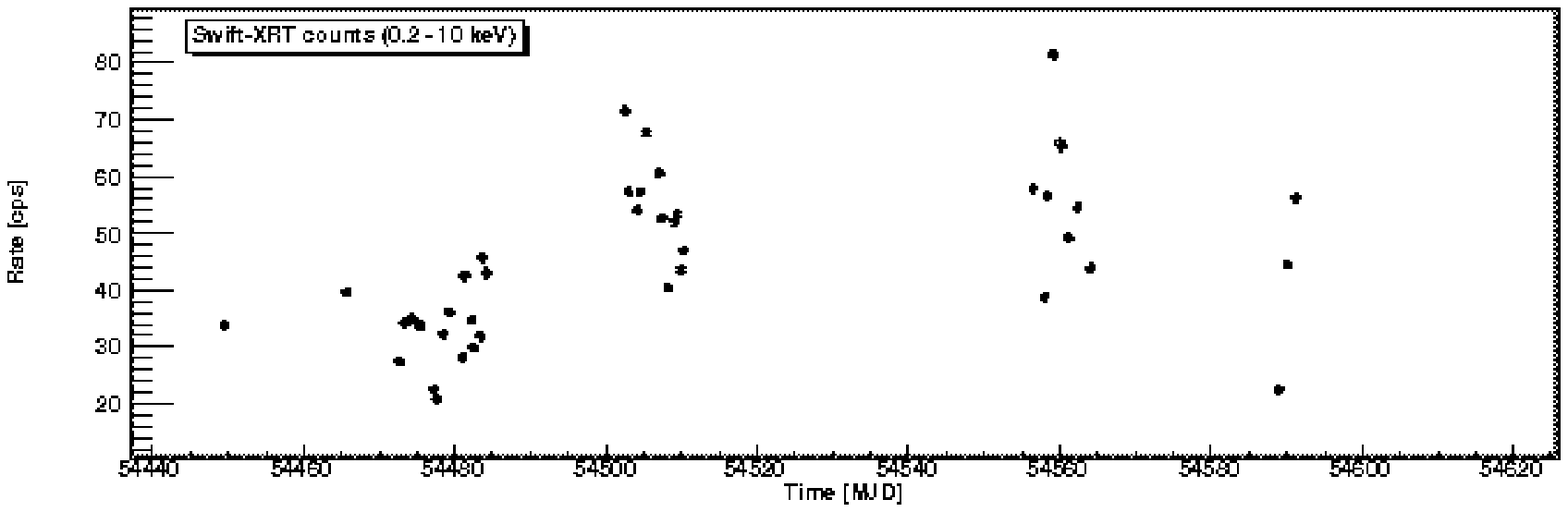}
  \includegraphics[width=4.5in]{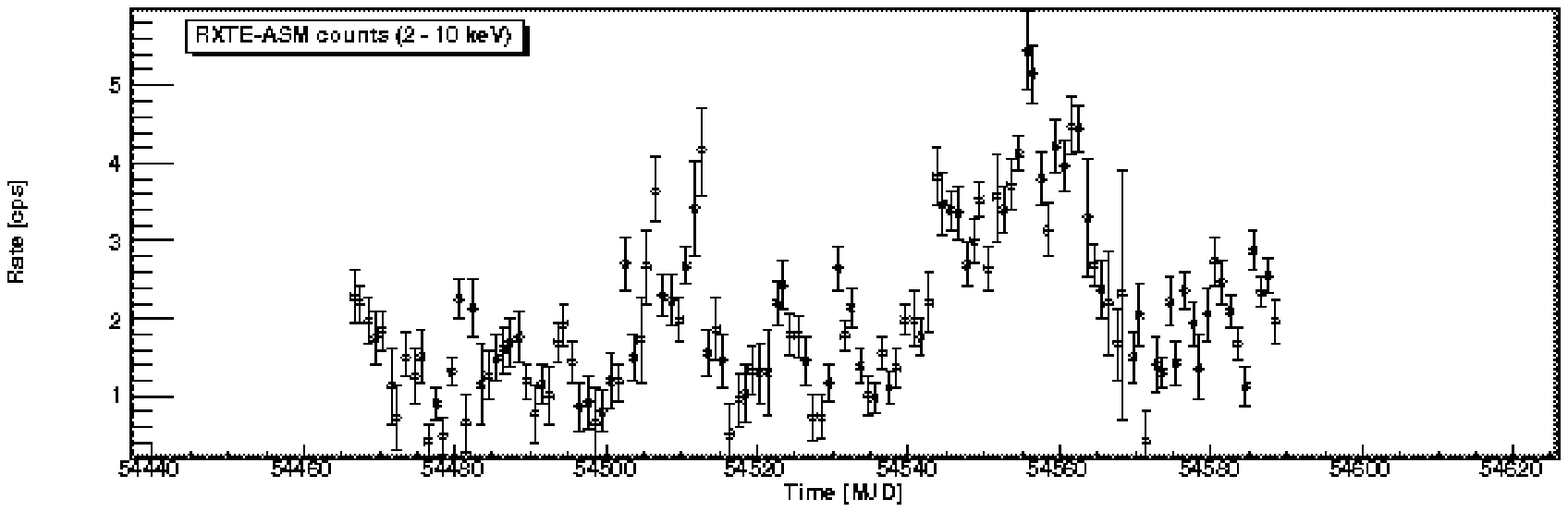}
  \includegraphics[width=4.5in]{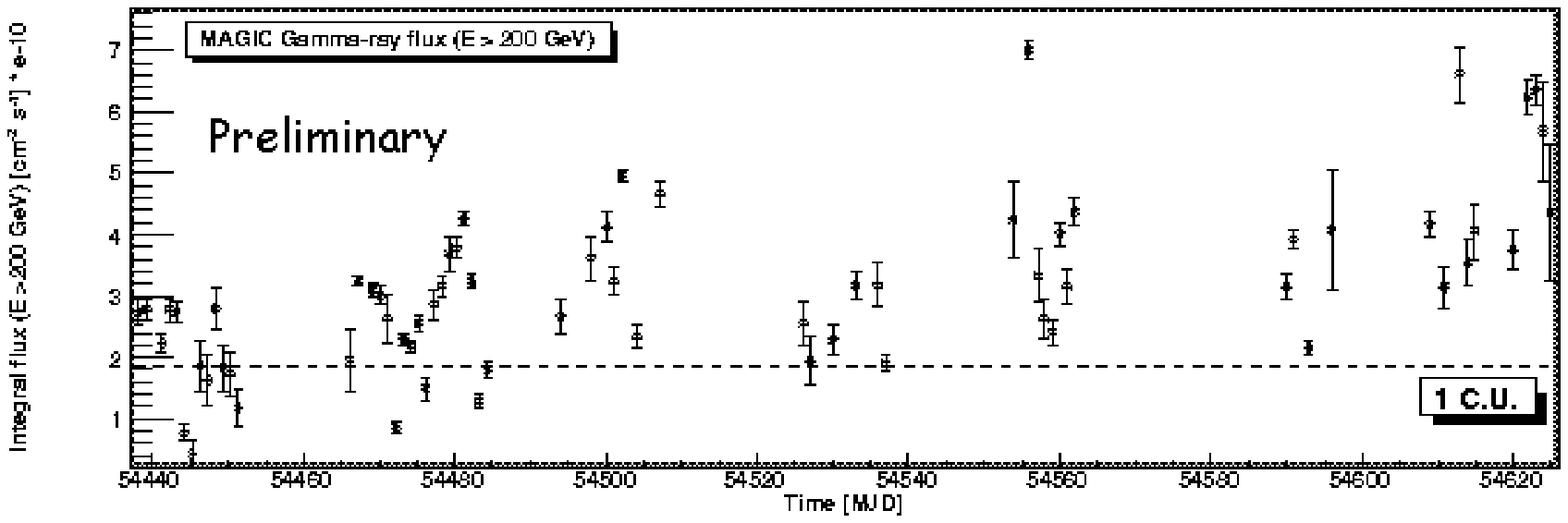}
  \caption{Upper panel: R-band optical lightcurve of Mkn 421 from the Tuorla Observatory along the MAGIC observation period. Middle-upper panel: soft X-ray count rates measured by Swift-XRT along the same period. Middle-lower panel: soft X-ray count rates measured by RXTE-ASM.
Lower panel: MAGIC VHE lightcurve above 200 GeV, for the 66 observation nights that passed quality cuts. MAGIC has detected the source clearly in all the nights; the integral flux level is lower than $2.0 \times 10^{-10}$ ($\sim 1\,  C.U.$, represented by the dashed horizontal line) only in a few nights. A maximum flux of $\sim 3.6$  C.U. has been observed on March 2008, the 30th.}
  \label{MWL_LC}
 \end{figure*}

\section{Magic Dataset}

MAGIC observed Mnk 421 in 81 nights since December 2007 until June 2008, which is the full visibility window for this source. Data were taken in ``wobble'' mode, originally suggested by Fomin \cite{1994APh.....2..151F}, which allows simultaneous signal and background measurement. The method avoids the need of devoting observing time to the measurement of background. This is of paramount importance here, as the source is variable and full time coverage of it's evolution is highly desirable. After quality cuts, $\approx 60$ hours of observations in 66 nights were available for further analysis. The zenith distance of the source ranged from $\sim 5^\circ$ at culmination to $46^\circ$. 

\subsection{Data analysis}

The data were analyzed with the standard MARS\footnote{MAGIC Analysis and Reconstruction Software} package.
The suppression of NSB residuals survived to the trigger selection and the calculation of image parameters \cite{1985ICRC....3..445H} were performed with the new timing techniques. 
The rejection of hadronic background and the estimation of the energy for the primary VHE $\gamma$-rays were performed by means of an implementation of the Random Forest classification method \cite{Breiman}.
Two new parameters were calculated, from the image and time parameters, for each event:  \emph{Hadronness} and \emph{Estimated Energy}.

The signal extraction was performed by applying cuts in the \emph{Size}, \emph{Hadronness} and \emph{Alpha} parameters. In particular the \emph{Size} cut, rejecting events with less than 150 photoelectrons of total charge, leads to an energy threshold $\sim 140\,$  GeV in the present analysis.
A total of $\sim 48 \times 10^3$ excess events from the selected 60 hours of observation were detected; the excess was statistically significant (above $5\sigma$ according to Li\&Ma statistics \cite{1983ApJ...272..317L}) in each energy bin from 90 GeV to 8 TeV. 

The whole analysis procedure was step by step validated on compatible datasets from observations of the Crab Nebula, widely regarded as the standard candle for VHE astronomy. The results were in agreement with the known properties of the source. Double cross-checks of the analysis gave compatible results.

\subsection{MAGIC VHE lightcurves}

\begin{figure}[!t]
  \centering
  \includegraphics[width=2.8in]{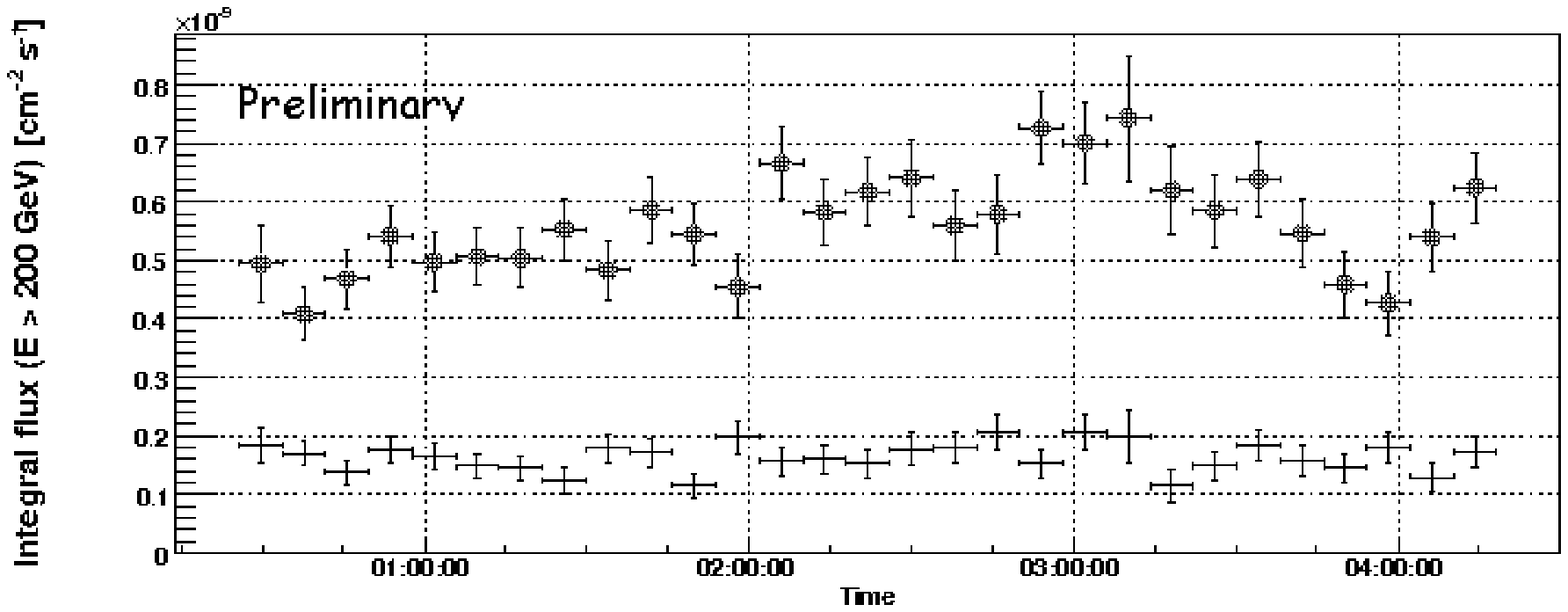}
\includegraphics[width=2.8in]{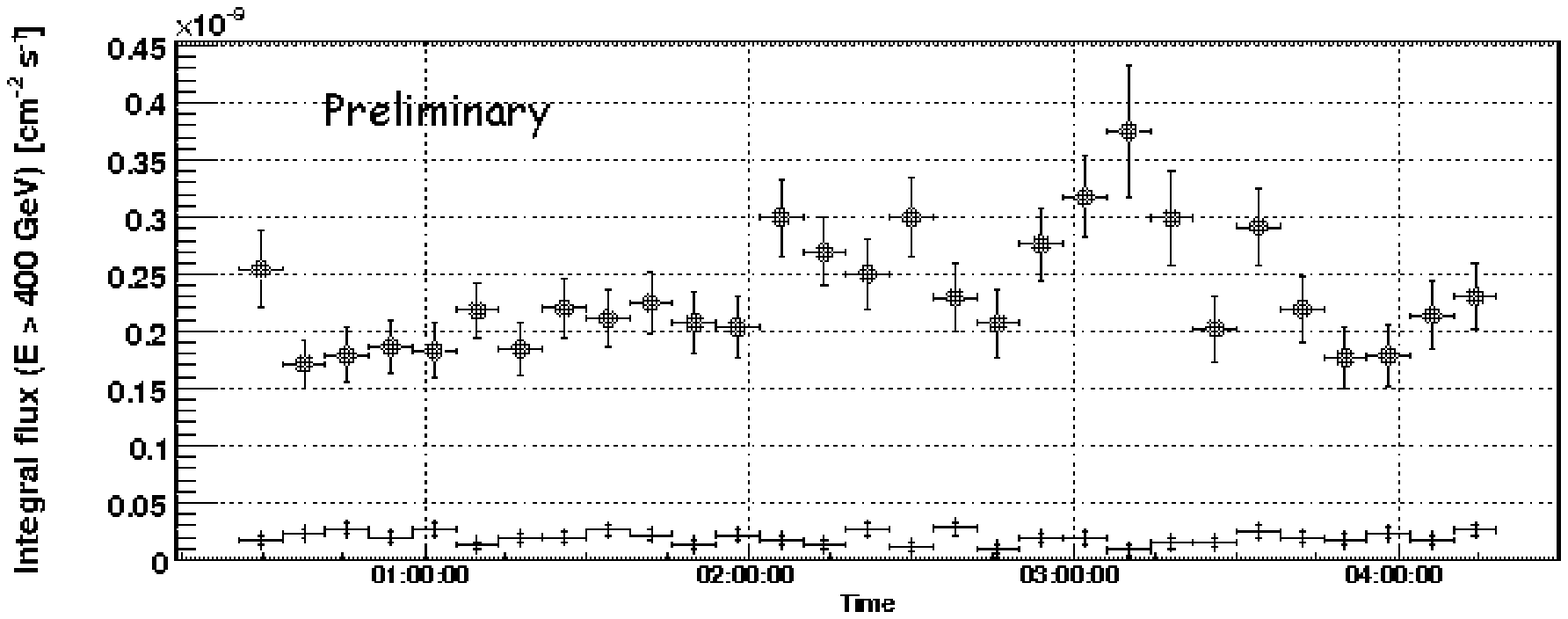}
  \caption{Mkn 421 VHE lightcurves in 8 min. time bins, from the observations taken on 2008, February the 6th.  Integral flux of excess (filled circles) and background (thin crosses) events is plotted. The Energy threshold is 200 GeV (upper panel) and 400 GeV (lower panel).}

  \label{LC_20080206}
 \end{figure}

\begin{figure}[!t]
  \centering
  \includegraphics[width=2.8in]{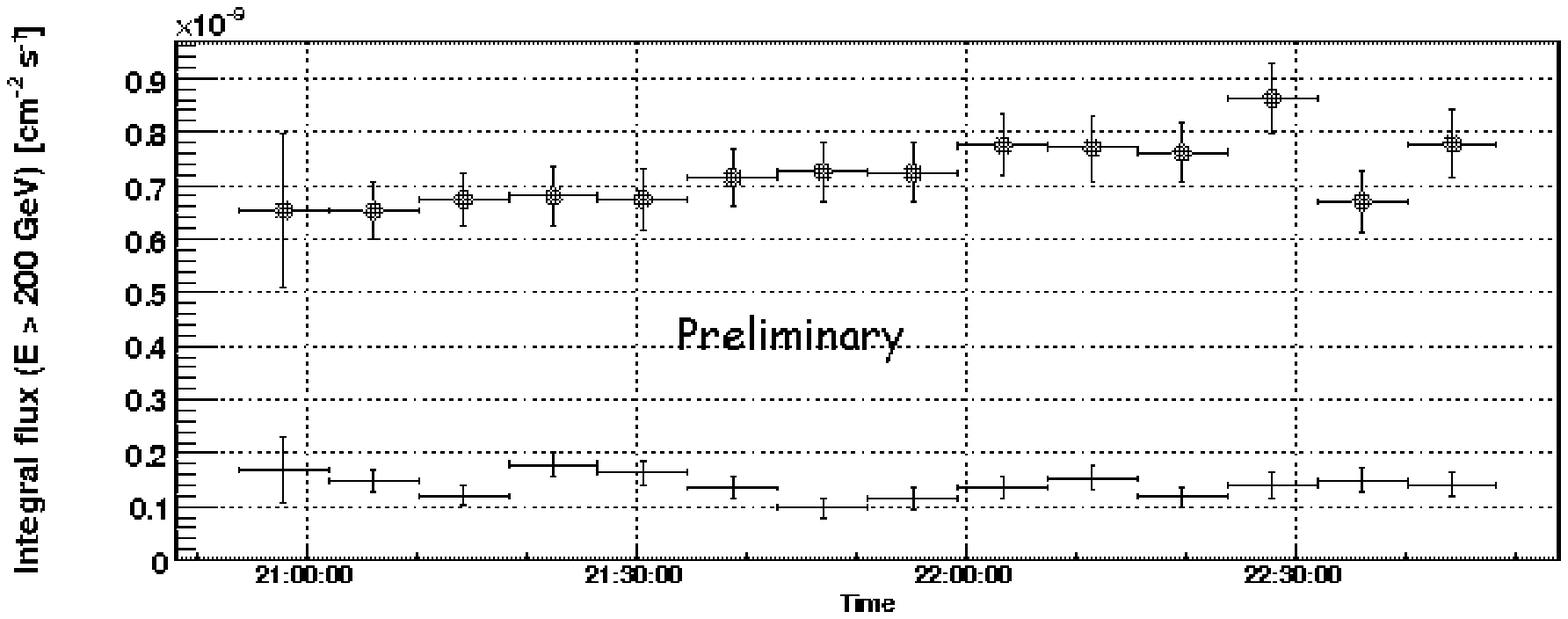}
\includegraphics[width=2.8in]{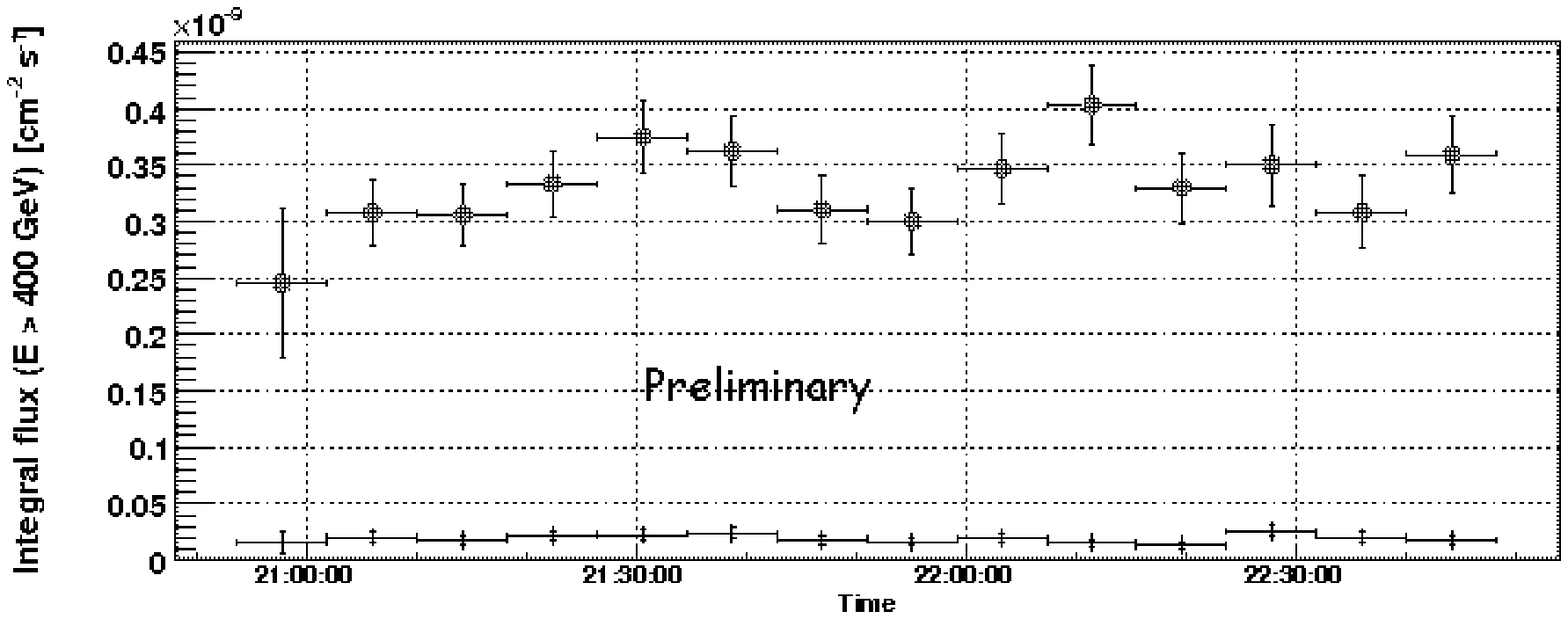}
  \caption{Mkn 421 VHE lightcurves in 8 min. time bins, from the observations taken on 2008, March the 30th. Integral flux of excess (filled circles) and background (thin crosses) events is plotted. The Energy threshold is 200 GeV (upper panel) and 400 GeV (lower panel).}

  \label{LC_20080331}
 \end{figure}

\begin{figure}[!t]
  \centering
\includegraphics[width=2.1in]{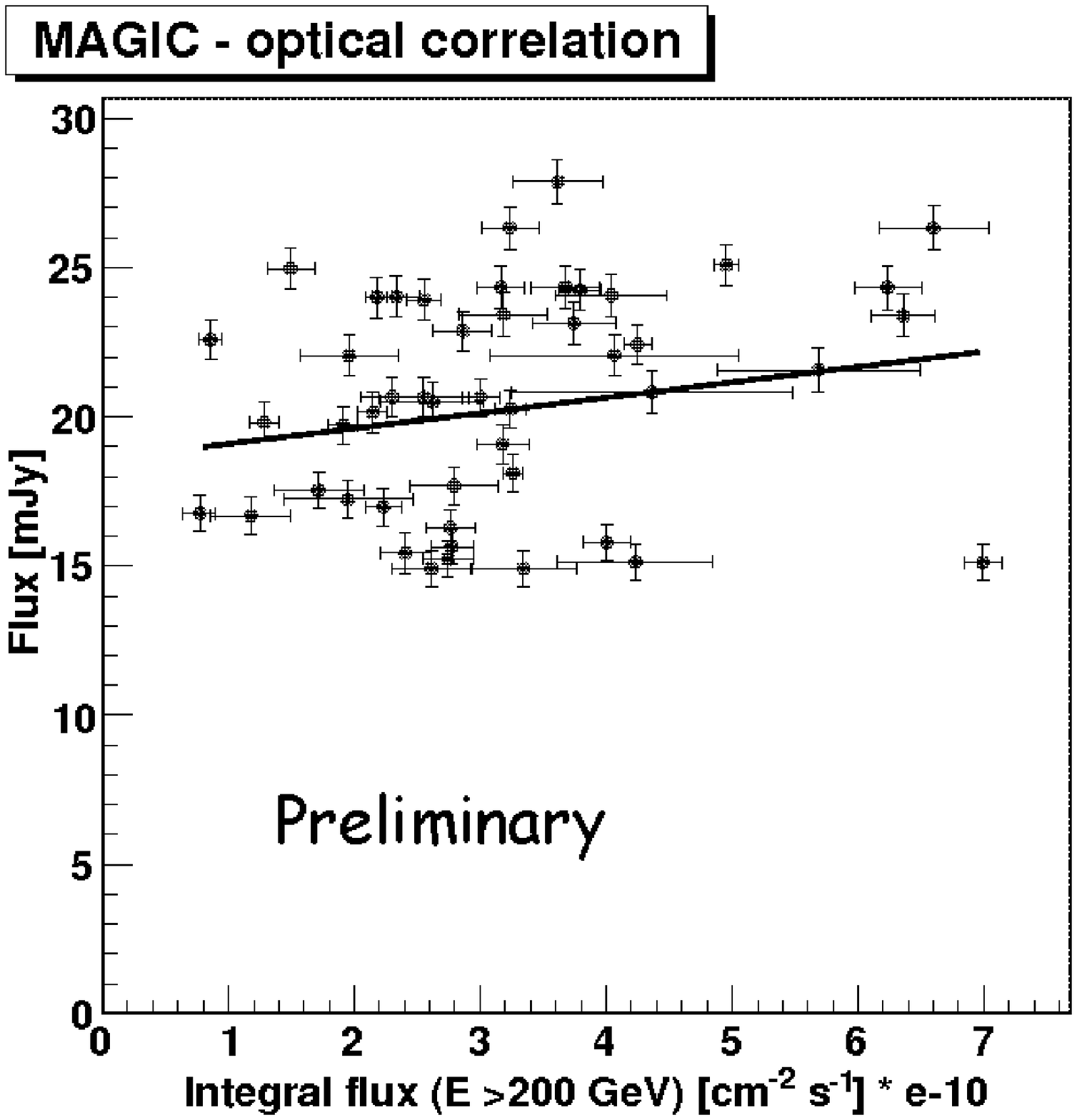} 
\includegraphics[width=2.1in]{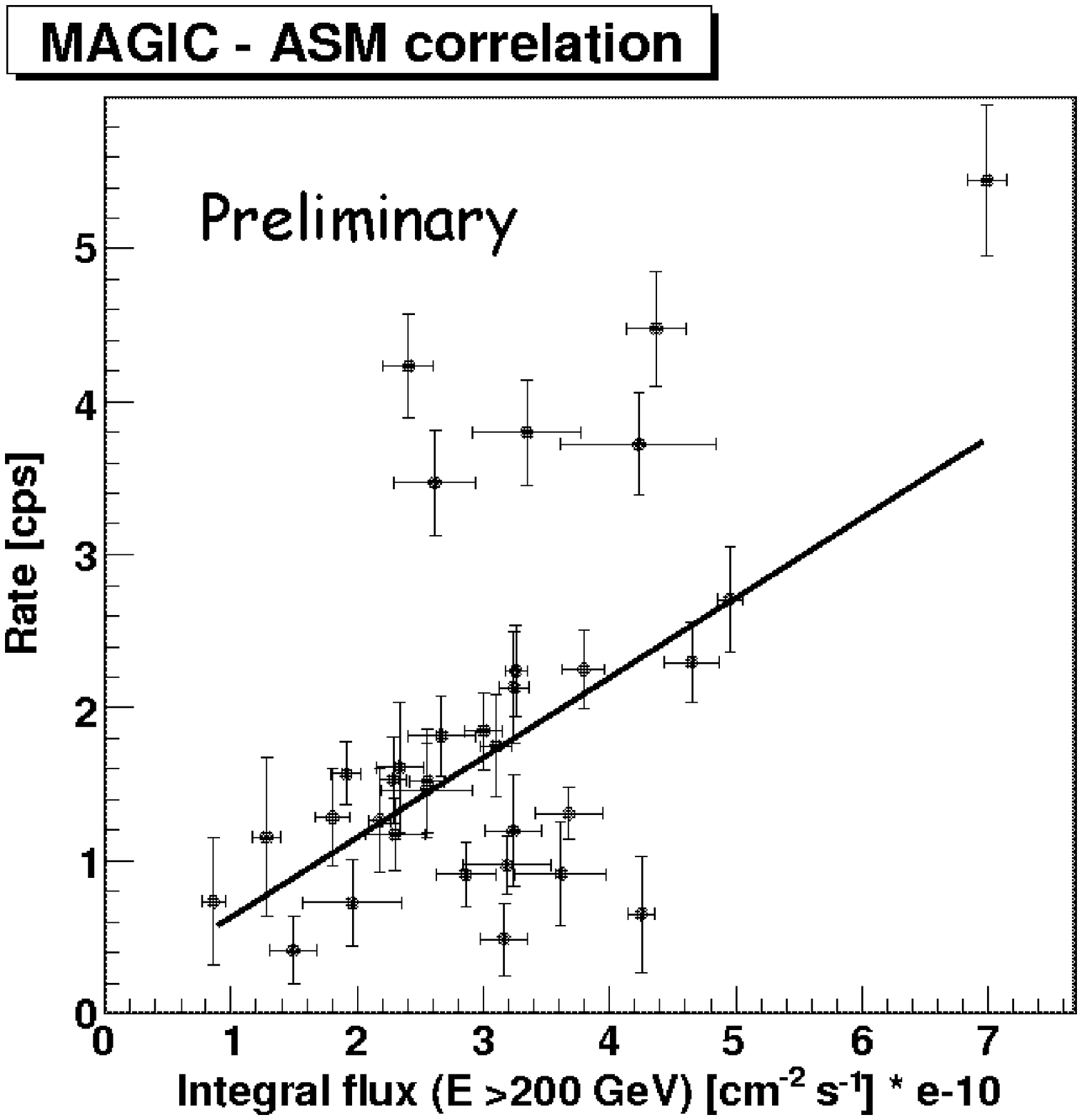} 
\includegraphics[width=2.1in]{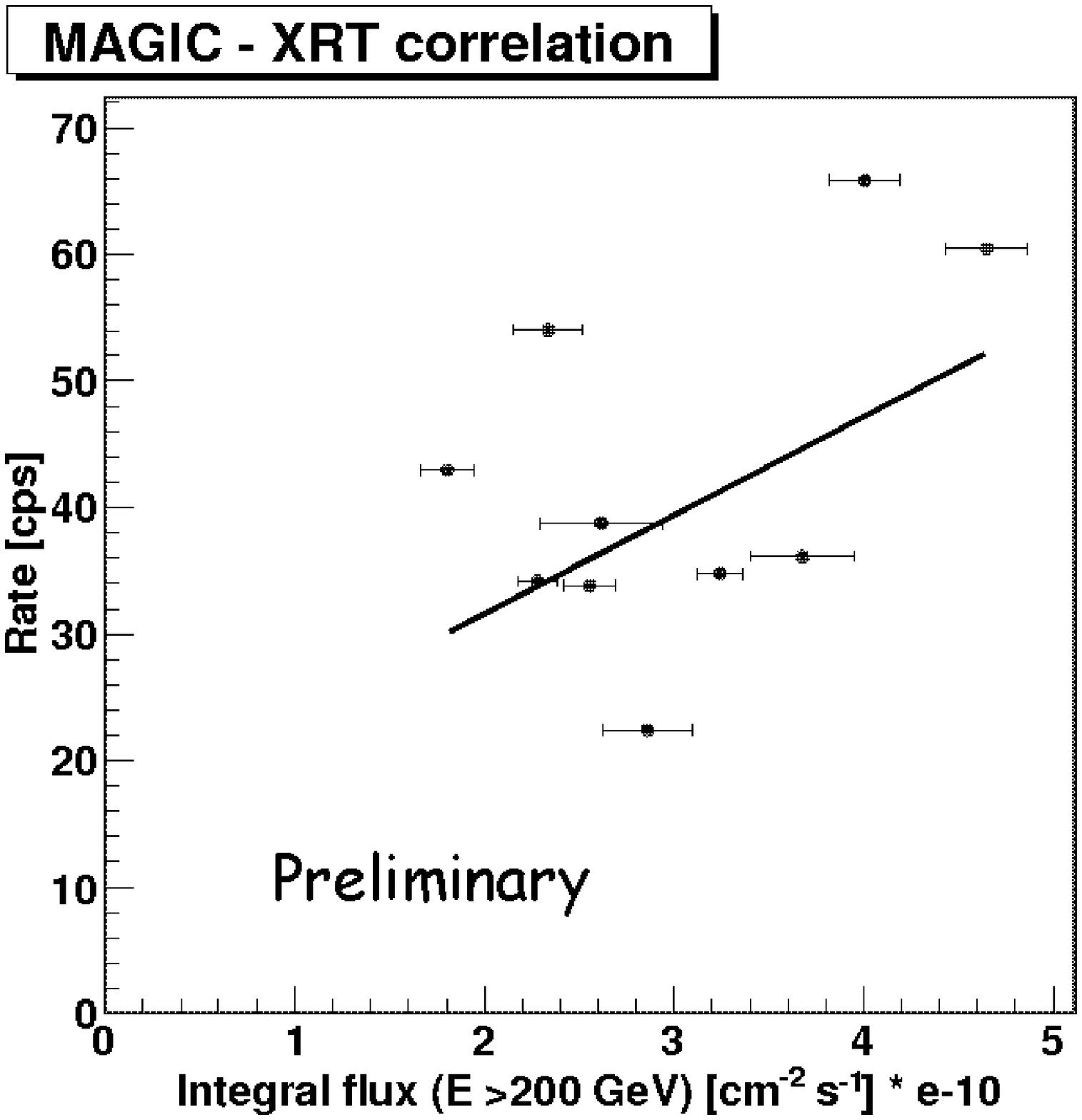} 

  \caption{Correlation plots for the MWL observations of Mkn 421. Upper panel: optical (R-band) vs. MAGIC VHE data. Within a time window of 0.5 days, 48 observations have been matched. The probability of independence of data is $3.88\%$ ($\rho=0.26$). Middle panel: RXTE-ASM X-ray data vs. MAGIC VHE data. Within a time window of 0.5 days, 33 couples of points were matched. The probability of independence of data is  $0.17\%$, $\rho=0.49$. Lower panel: correlation plot for the 10 tightly simultaneous observations from MAGIC and Swift-XRT. The probability of independence is $18.2\%$, $\rho=0.32$.}
\label{correlations}
\end{figure}

The night-averaged integral flux above 200 GeV was calculated for the 66 nights with datasets surviving the quality cuts. The lightcurve is shown in figure \ref{MWL_LC}.
It's worth noticing that even if Mkn 421 is believed to emit a low VHE flux
baseline \cite{1996ApJ...460..644S}, the flux was seldom below the Crab level ($F_{E>200\, GeV} = 1.96\pm 0.05_{stat} \times 10^{-10} cm^{-2} s^{-1}$ \cite{2008ApJ...674.1037A}) all along the period,
confirming an intense and persistent active state. The maximum observed flux ($F_{E>200\, GeV} = 6.99\pm 0.15_{stat} \times 10^{-10} cm^{-2} s^{-1}$) was on March 2008, the 30th.

In particular, on February the 6th and March the 30th long observations could be taken of very high flux states, under good weather conditions.
Lightcurves in 8 min. time bins are shown for these nights (figures \ref{LC_20080206} and \ref{LC_20080331} respectively) with a softer ($E > 200\, GeV$, upper panels) and harder ($E> 400\, GeV$, lower panels) energy threshhold. Flux variability with doubling times down to  16 min can be seen in the first night, while in the second night no rapid evolution is visible.

\section{MWL data}

For the multiwavelenght study optical (R-band) and soft-X ($0.2-10\,keV$) data have been collected.

\subsection{Optical Data}

The Tuorla Observatory constantly monitors the MAGIC VHE (known or potential) sources, by means of the $35\, $cm remotely operated KVA optical telescope also located at Roque de los Muchachos and of a $103\,$ cm telescope located at Tuorla, Finland. Along the period of MAGIC observations, 117 flux measurement of Mkn 421 were performed in the Johnson $R$ band. The light contribution from the host galaxy and nearby companion galaxy ($F_{h+cg} =8.07\pm 0.47\,$ mJy \cite{2007A&A...475..199N}) has been subtracted from the measured fluxes.
The optical lightcurve is reported in the upper panel of figure \ref{MWL_LC}. 

\subsection{X-ray data from  Swift-XRT}
The X-Ray Telescope (XRT) onboard of the NASA Swift mission has observed Mkn 421 43
times along the period, mainly with short ($1-2$ ks) exposures. Swift-XRT data were reduced using the xrtpipeline software distributed with the heasoft 6.3.2 package  by the NASA High Energy Astrophysics Archive Research Center (HEASARC).
The observed count rates in the $0.2-10\,$  keV band are reported in the second panel of figure \ref{MWL_LC}.

\subsection{X-ray data from RXTE-ASM}

Although with worse sensitivity and precision w.r.t. Swift-XRT the All-Sky Monitor onboard of the Rossi X-ray Timing Explorer is still able to set one point per day from Mkn 421.

Mkn 421 ASM data are taken from the results provided by the ASM/RXTE teams at MIT and at the RXTE SOF and GOF at NASA's GSFC.
The count rates observed in the $2-10\,$ keV band are shown in the third panel of figure \ref{MWL_LC}.

\section{Correlations}

VHE data have been correlated to the available data in the other bands.
 Due to the general lack of tightly simultaneous observations, points in the different bands taken within a  time window of  0.5 days have been accepted as matching. This is quite acceptable for the optical, where variations can be assumed small on these timescale. As far as RXTE-ASM measurements are concerned, these come from a 24-hours average, so the assumption is fully valid. For the correlation between Swift-XRT and MAGIC observations, the assumption is no longer valid, so the correlation has been restricted to the 10 tightly simultaneous observation slots alone.
The results are that optical and VHE (figure \ref{correlations}, upper panel) are partially correlated (Spearman's rank \cite{Hollander} $\rho=0.26$, probability that the datasets are independent $P(null)=3.88\%$)
Stronger correlation is found between RXTE-ASM and MAGIC data (figure \ref{correlations}, middle  panel): $\rho=0.49$, $P(null)=0.17\%$.
The result of correlation between Swift-XRT and MAGIC data is clearly  suffering from low statistics: $\rho=0.32$, $P(null)=18.2\%$. No dramatic improvement is seen by relaxing the request on tight simultaneity, in spite of the greater statistics. This issue cannot be satisfactorily explained here, but could be due to the variation of the source state between time-separated measurements.  

\section{Conclusions}

The MAGIC Telescope has extensively monitored the TeV HBL Mkn 421 since December 2007 until June 2008. 

The observed X-ray vs. VHE correlation is in full agreement with previous observations (e.g. \cite{2008ApJ...677..906F}) and fits naturally the SSC scenario, where the same population of relativistic electrons is involved in both  the synchrotron and the Inverse Compton emission, although hadronic models cannot be ruled out. 
The weaker correlation between optical and VHE state can be regarded as due to a contribution to the optical flux from regions of the jet that are different from the one producing VHE $\gamma$-rays.

\section{ ACKNOWLEDGMENT}
  We would like to thank the Instituto de Astrofisica de Canarias for the excellent working conditions at the Observatorio del Roque de los Muchachos in La Palma. The support of the German BMBF and MPG, the Italian INFN and Spanish MICINN are gratefully acknowledged.


\begin{thebibliography}{99}
\bibitem{2007ApJ...669..862A} Albert, J., et al.\ 2007, ApJ, 669, 862
\bibitem{2008ApJ...674.1037A} Albert, J., et al.\ 2008, ApJ, 674, 1037
\bibitem{Breiman} Breiman, L., Machine Learning, 45, 4
\bibitem{2005ICRC....5..359C} Cortina, J., \& et al.\ 2005, International Cosmic Ray Conference, 5, 359
\bibitem{1991trcb.book.....D} de Vaucouleurs, G., de Vaucouleurs, A., Corwin, H.~G., Jr., Buta, R.~J., Paturel, G., \& Fouque, P.\ 1991, Volume 1-3, XII, 2069 pp.~7 figs..~ Springer-Verlag Berlin Heidelberg New York,
\bibitem{1994APh.....2..151F} Fomin, V.~P., Fennell, S., Lamb, R.~C., Lewis, D.~A., Punch, M., \& Weekes, T.~C.\ 1994, Astroparticle Physics, 2, 151
\bibitem{2008ApJ...677..906F} Fossati, G., et al.\ 2008, ApJ, 677, 906
\bibitem{1996Natur.383..319G} Gaidos, J.~A., et al.\ 1996, \emph{Nature}, 383, 319
\bibitem{1985ICRC....3..445H} Hillas, A.~M.\ 1985, International Cosmic Ray Conference, 3, 445
\bibitem{Hollander} M. Hollander, D.A. Wolfe, \emph{Nonparametric statistical methods}, Wiley (1973)
\bibitem{2004ApJ...601..151K} Krawczynski, H., et al.\ 2004, ApJ, 601, 151
\bibitem{2006ApJ...651..113K} Kusunose, M., \& Takahara, F.\ 2006, ApJ, 651, 113
\bibitem{1983ApJ...272..317L} Li, T.-P., \& Ma, Y.-Q.\ 1983, ApJ, 272, 317 
\bibitem{1993A&A...269...67M} Mannheim, K.\ 1993, A\&A, 269, 67
\bibitem{1992ApJ...397L...5M} Maraschi, L., Ghisellini, G., \& Celotti, A.\ 1992, ApJL, 397, L5
\bibitem{1997jena.confE.279M} Mastichiadis, A., \& Kirk, J.~G.\ 1997, Joint European and National Astronomical Meeting,
\bibitem{2003APh....18..593M} M{\"u}cke, A., Protheroe, R.~J., Engel, R., Rachen, J.~P., \& Stanev, T.\ 2003, Astroparticle Physics, 18, 593
\bibitem{2007A&A...475..199N} Nilsson, K., Pasanen, M., Takalo, L.~O., Lindfors, E., Berdyugin, A., Ciprini, S., \& Pforr, J.\ 2007, A\&A, 475, 199 
\bibitem{1992Natur.358..477P} Punch, M., et al.\ 1992, \emph{Nature}, 358, 477
\bibitem{1996ApJ...460..644S} Schubnell, M.~S., et al.\ 1996, ApJ, 460, 644
\bibitem{1998ApJ...509..608T} Tavecchio, F., Maraschi, L., \& Ghisellini, G.\ 1998, ApJ, 509, 608
\bibitem{2008ICRC....3.1393T} Tescaro, D., Bartko, H., Galante, N., \& et al.\ 2008, International Cosmic Ray Conference, 3, 1393
\bibitem{1995PASP..107..803U} Urry, C.~M., \& Padovani, P.\ 1995, \emph{PASP}, 107, 803




















  \end{thebibliography}
\end{document}